# Surface diffusion control of the photocatalytic oxidation in air/TiO₂ heterogeneous reactors

R. Tsekov,[1] E. Evstatieva[2] and P.G. Smirniotis[3]
[1]Department of Physical Chemistry, University of Sofia, 1164 Sofia, Bulgaria
[2]Department of Water Technology, Karlsruhe Research Center, 76021 Karlsruhe, Germany
[3]Department of Chemical Engineering, University of Cincinnati, Cincinnati, OH 45221, USA

The diffusion of superoxide radical anions on the surface of TiO₂ catalysts is theoretically considered as an important step in the kinetics of photocatalytic oxidation of toxic pollutants. A detailed analysis is performed to discriminate the effects of rotation, anion and adsorption bonds vibrations on the diffusion coefficient. A resonant dependence of the diffusivity on the lattice parameters of the TiO₂ surface is discovered showing that the most rapid diffusion takes place when the lattice parameters are twice larger than the bond length of the superoxide radical anions. Whereas the rotation and vibrations normal to the catalyst surface are important, the anion bond vibrations do not affect the diffusivity due to their low amplitudes as compared to the lattice parameters.

Photocatalytic oxidation of organic compounds is an alternative to the conventional methods for removal of toxic pollutants from air and water [1, 2]. Among the applied semiconductor photocatalysts [3] the most promising one is the titanium dioxide [4] because of its high photocatalytic activity, resistance and low costs. It is well known that valence band holes and conduction band electrons are generated by illumination of TiO₂ samples with light energy greater than the band gap (3.2 eV) $TiO_2 \xrightleftharpoons{h\nu} h^+ + e^-$. Here a possible recombination of the generated carriers with an emission of heat is also described. The holes $h^+$ and electrons $e^-$ act on the TiO₂ surface as powerful oxidizing and reducing agents, respectively. The holes oxidize water molecules, adsorbed at the TiO₂ surface from the air, to hydroxyl radicals $H_2O + h^+ \to {}^\bullet OH + H^+$ while the electrons react with oxygen molecules adsorbed on the Ti(III)-sites reducing them to superoxide radical anions $O_2 + e^- \to {}^\bullet O_2^-$. The hydroxyl and superoxide radicals are responsible for the oxidizing steps in photocatalysis [5]. The generally agreed mechanism involves an initial attack of the organic molecule by a hydroxyl radical followed by a reaction of the resulting radical with superoxide radicals. The problem is that several ${}^\bullet O_2^-$ are required for complete mineralization of a toxic molecule. Hence, the oxidation process can be limited by the mobility of the superoxide radical anions on the TiO₂ surface and, for this reason, the description of their surface diffusion is very important for the kinetic modeling of the air/TiO₂ reactors [6].

The diffusion of particles in solids and on their surfaces is of current interest both for the pure science and technology. The diffusion coefficient of molecules in solids exhibits non-monotonous dependence on the molecular size, a phenomenon called resonant diffusion [7]. An interfacial example of this phenomenon represents the Brownian motion of rhenium dimers on tungsten [8] and ruthenium clusters on rhenium [9]. The resonant diffusion is theoretically described via molecular dynamics simulations [10, 11] and stochastic modeling [12-14]. Some general treatments of the diffusion in modulated structures are also developed [15-20]. The problem of diffusion on solid surfaces is equivalent to analysis of the effect of the thermal phonons in the solid on the behavior of the adsorbed molecule. A rigorous consideration of the stochastic dynamics of a particle coupled with a harmonic bath is possible [16, 19]. As a result, one yields a dynamic equation with a position dependent friction and a random force, which are coupled by the fluctuation-dissipation theorem. The comparison of the theory with experiments for diffusion of rhenium atoms and dimers on tungsten surfaces is excellent [13]. The effects of dimmer rotation [21] and vibration [22] on the barrier crossing are also taken into account. The present study aims to apply this model for description of the diffusion of superoxide radical anions on $TiO_2$ surface in order to discover the optimal conditions for photocatalytic oxidation of pollutants in air/TiO2 reactors.

Let us consider first the model of a rigid superoxide anion with a fixed distance between the two oxygen atoms. The potential $\phi$ is the fundamental measure of the interaction between an oxygen atom and the $TiO_2$ surface, which can be rigorously calculated by *ab initio* methods. However, since the catalysts surfaces are not well defined and covered by various equilibrium and non-equilibrium interfacial structures, it is necessary to introduce an effective model for the potential $\phi$. Since the latter is a periodic function, which possesses the symmetric properties of the local $TiO_2$ lattice, $\phi$ can be naturally expanded in a Fourier series. Assuming that there is a dominate term in this series, hereafter the following model well known from the literature [12,21] will be employed

$$\phi = A\cos(2\pi x/a) + A\cos(2\pi y/b) \tag{1}$$

where $x$ and $y$ are the coordinates of the oxygen atom and $A$ is the magnitude of the interaction. By variation of the parameters $a$ and $b$ one is able to model different lattices on the $TiO_2$ surface. The basic advantage of the decoupled cosine potential (1) is that the corresponding forces split to independent components, i.e. the forces acting of $x$- and $y$-axes depend on $x$ and $y$, respectively. It is easy now to write the potential energy of the superoxide radical anion in the form

$$U = \phi_1 + \phi_2 = 2A\cos(\pi\cos\varphi L/a)\cos(2\pi X/a) + 2A\cos(\pi\sin\varphi L/b)\cos(2\pi Y/b) \tag{2}$$

where $X$ and $Y$ are the coordinates of the mass center of the anion, $\varphi$ is the angle between the anion and $x$-axis and $L$ is the anion length. Another characteristic important for the Brownian motion of the anions is the friction coefficient $B$. It is possible starting from classical mechanics to derive the following expression [17] for the friction coefficient acting on the $x$-axis

$$4\pi\rho c^3 B_X = (\partial^2\phi_1/\partial x_1^2)^2 + (\partial^2\phi_2/\partial x_2^2)^2 = (16\pi^4 A^2/a^4)[1+\cos(2\pi\cos\varphi L/a)\cos(4\pi X)] \quad (3)$$

where $\rho$ and $c$ are the mass density and sound velocity in $TiO_2$, respectively. The friction coefficient on the $y$-axis can be calculated analogically.

Due to the simplicity of the potential (1), one can directly apply the well-known formula

$$D_X = \frac{a^2}{\beta \int_0^a B_X \exp(\beta U)dX \int_0^a \exp(-\beta U)dX} \quad (4)$$

for calculation of the diffusion constant, which is rigorously derived for the case of one dimensional motion of a particle in a periodic potential [15]. Here $\beta = (kT)^{-1}$ is the inverse thermodynamic temperature. Equation (4) shows a relative asymmetric effect of the friction coefficient to diffusion. Since the exponential term in the first integral is mainly determined by the maximum of $U$, the friction on the top of the potential barriers is more essential for the value of $D_X$, while the value of $B_X$ in the potential wells is of second order of importance. Substituting the expressions (2) and (3) in eq 4 one is able to calculate analytically the diffusion coefficient of the superoxide radical anion

$$D_X = \frac{\rho c^3 a^4}{4\pi^3 AI_0[2\beta A\cos(\pi\cos\varphi L/a)]} \times \frac{\cos(\pi\cos\varphi L/a)}{2\beta A\cos^3(\pi\cos\varphi L/a)I_0[2\beta A\cos(\pi\cos\varphi L/a)] - \cos(2\pi\cos\varphi L/a)I_1[2\beta A\cos(\pi\cos\varphi L/a)]} \quad (5)$$

where $I_0$ and $I_1$ are the modified Bessel functions of zero and first order, respectively. In Fig. 1 the dependence of the diffusion coefficient from eq 5 is plotted as a function of the angle between the anion and $x$-axis and the corresponding lattice parameter $a$. The specific material constants are equal to $\rho = 4260$ kg/m³, $c = 6900$ m/s, $T = 350$ K and $L = 1.35$ Å. The value $2A = 5\times10^{-20}$ J is estimated as the adsorption heat of oxygen on $TiO_2$ surface [23]. As seen from

Fig. 1 there is a strong resonant dependence of the diffusion coefficient, which exhibits a maximum when the lattice parameter $a$ is twice larger than the $x$-projection of the anion. In this case the activation energy is zero.

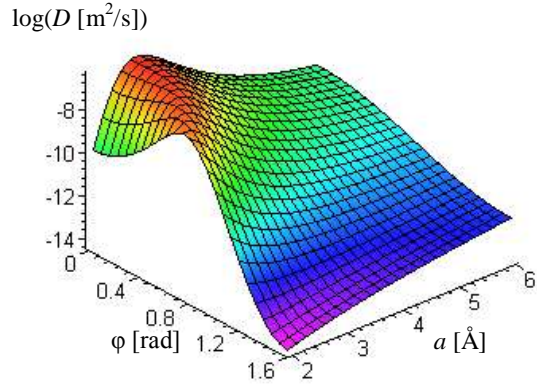

**Fig. 1.** Dependence of the diffusion coefficient of a superoxide radical anion on the angle $\varphi$ and the lattice parameter $a$

A further problem is the influence of the rotation of the adsorbed superoxide radical anions on the resonant dependence of the diffusion coefficient. The complete investigation of the dynamics of a rotating anion requires a stochastic description of both the translation and rotation. A basic approximation here follows from the difference between the characteristic timescales of these two processes [21]. In this case one can apply adiabatic separation of the dynamics to slow (translation) and quick (rotation) motions. This means, that from the viewpoint of translation the rotation is always at equilibrium. Thus, using the potential energy of the anion from eq 2 one can calculate the rotational integral of state

$$Z_R = \int_0^{2\pi} \exp(-\beta U) d\varphi \qquad (6)$$

The knowledge of $Z_R$ allows calculation of the rotational free energy of the anion $F_R = -\beta^{-1} \ln Z_R$, which plays the role of an average potential for translation. Unfortunately, the function $F_R$ is not a sum on independent components corresponding to the two axes of motion. For this reason we will employ further a second adiabatic procedure to take into account the influence of the motion on $Y$ to the motion on $X$. The corresponding partial translation integral of state acquires the form

$$Z_T = \frac{1}{b}\int_0^b \exp(-\beta F_R) dY = \int_0^{2\pi} \exp[-2\beta A \cos(\pi \cos\varphi L/a)\cos(2\pi X/a)] I_0[2\beta A \cos(\pi \sin\varphi L/b)] d\varphi$$

By means of $Z_T$ one can calculate the conditional free energy $F_T = -\beta^{-1} \ln Z_T$ for the anion motion on the $x$-axis. Of course, $F_T$ is a periodic function of $X$ and, consequently, one can apply once again the formula (4) to calculate the diffusion coefficient of the rotating anion

$$D_X(F_T) = \frac{\rho c^3 a^6}{8\pi^3 \beta A^2 \int_0^a \exp(\beta F_T) dX \int_0^a \exp(-\beta F_T) dX} \qquad (7)$$

Note that in eq 7 the friction coefficient is replaced by its value at the maximum of the potential. The analytical integration of this formula is complex and time consuming and, for this reason, numerical calculations will be employed, where the values of the specific constants are fixed as mentioned above. The two lattice parameters $a$ and $b$ are varied in the process of calculations and the result is presented in Fig. 2. As seen there is a pronounced maximum of the diffusion constant at $a = b = 2L$. Another interesting feature is the asymmetric effect of the two lattice parameters: the effect of the lattice parameter on the complementary axis $b$ is weaker as compared to the effect of the parameter $a$.

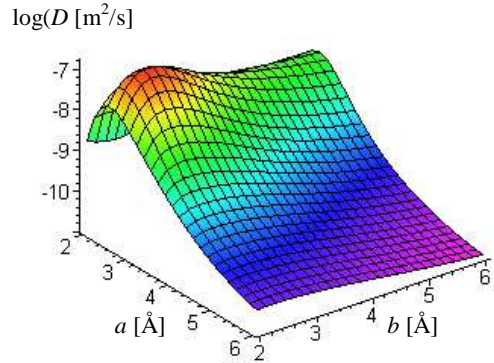

**Fig. 2.** Dependence of the diffusion coefficient of a rotating superoxide radical anion on the lattice parameters $a$ and $b$ at the equilibrium distance from the TiO$_2$ surface

In the present part, a method based on eq 4 is developed for calculation of the diffusion coefficient of a vibrating anion on the TiO$_2$ surfaces. In general, there are two kinds of vibrations relevant to the process of diffusion: internal vibrations of the anion bond and vibrations of the anion normal to the TiO$_2$. According to eq (2), the potential energy of a superoxide radical anion with arbitrary length is

$$U = 2A\cos(\pi \cos\varphi L\xi / a)\cos(2\pi X / a) + 2A\cos(\pi \sin\varphi L\xi / b)\cos(2\pi Y / b) \qquad (8)$$

where $\xi$ is the relative bond length as compared to the equilibrium value $L$. As was mentioned above, owing to the difference in the characteristic timescales of the anion translation and vibration, it is possible to apply an adiabatic separation of the slow and quick variables. The latter should be considered in thermodynamic equilibrium. Therefore, the effect of the bond vibrations can be accounted for by the vibration partition function

$$Z_V = \int_0^\infty \int_0^{2\pi} \exp[-2\beta A \cos(\pi \cos \varphi L \xi / a) \cos(2\pi X / a)] I_0 [2\beta A \cos(\pi \sin \varphi L \xi / b)] \exp(-\beta V) d\varphi d\xi \quad (9)$$

where $V$ is the internal potential energy of the anion. A convenient model for $V$ is the harmonic potential $V = 2\mu(\pi \omega L)^2 (\xi - 1)^2$, where $\mu = 8$ g/mol is the reduced mass of the anion and $\omega = 1090$ cm$^{-1}$ is the frequency of the superoxide anion vibrations [24]. Using this expression one can calculate $Z_V$ from eq 9 and the conditional free energy of the vibrating anion $F_V = -\beta^{-1} \ln Z_V$, which plays the role of an average potential acting on the anion translation. Hence, applying the formula (4) with a constant friction term leads to $D_X(F_V)$. Thus the calculated diffusion coefficient does not differ from the result in Fig. 2. This means that the interaction between the two oxygen atoms is very strong and the small bond vibrations do not affect the diffusion coefficient of the superoxide radical anion. One could argue that the value of $\omega$ for an adsorbed anion is lower than the used value from the gas phase. Further calculations show, however, that the bond vibrations become important if $\omega$ is bellow 200 cm$^{-1}$. Such an unrealistically low frequency could be observed only if the superoxide radical anion is in an exited state [25].

Other possible vibrations are those normal on the TiO$_2$ surface along the adsorption bond. In this case the potential energy of the superoxide radical anion acquires the form

$$U = -W \cos(\pi \cos \varphi L / a) \cos(2\pi X / a) - W \cos(\pi \sin \varphi L / b) \cos(2\pi Y / b) \quad (10)$$

where $W$ is the potential energy of the adsorption bond depending on the $z$-coordinate. If the interaction between an oxygen atom and the atoms from the solid obeys the Lenard-Jones potential, the adsorption van der Waals potential takes the form

$$W = A(1/\zeta^9 - 3/\zeta^3) \quad (11)$$

where $\zeta$ is the relative distance between the TiO$_2$ surface and the mass center of the superoxide radical anion as compared to the equilibrium distance. Obviously, $W$ exhibits a minimum at $\zeta = 1$, which is deep $2A$. Using eq 11, one can calculate the vibration partition function for the present case as

$$Z_W = \int_0^\infty \int_0^{2\pi} \exp[\beta W \cos(\pi\cos\varphi L/a)\cos(2\pi X/a)] I_0[-\beta W \cos(\pi\sin\varphi L/b)]\exp(-\beta W)d\varphi d\zeta \quad (12)$$

The internal anion vibrations are not considered in eq 12 since, according to the results above, they are not important due to the low amplitude of the O-O bond deviations. The free energy $F_W = -\beta^{-1}\ln Z_W$ for the translation follows directly from eq 12 and substituted in eq 7 it determines the diffusion coefficient of the superoxide anion. The numerical calculated $D_X(F_W)$ is presented in Fig. 3 as a function of the lattice parameters $a$ and $b$. As seen the normal vibration dramatically increases the diffusivity of the anions. The only exception is the case of the resonant maximum. Since there the activation energy is zero, it is not affected by the vibrations.

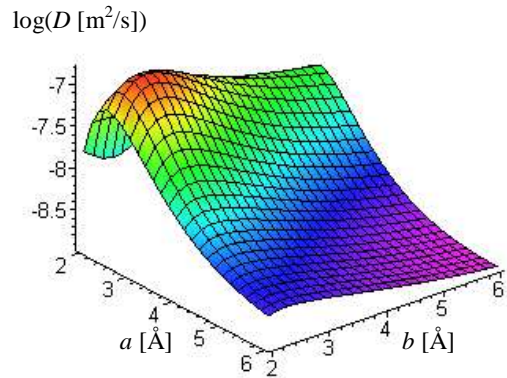

**Fig. 3.** Dependence of the diffusion coefficient of a rotating superoxide radical anion on the lattice parameters $a$ and $b$ with account for the vibrations normal to the TiO$_2$ surface

There are several effects important for the diffusion of superoxide radical anions on the TiO$_2$ surface. While the rotation and vibrations normal to the catalyst surface are essential, the anion bond vibrations do not affect the diffusivity. The dependence of the diffusion coefficient on the lattice parameters of the TiO$_2$ surface is strongly resonant and exhibits a pronounced maximum at lattice constants twice larger than the bond length of the superoxide radical anion. Hence, to increase the surface diffusion and thus the oxidation rate it is important to design the TiO$_2$ surface during the catalysts engineering in order to fit as well as possible the resonant condition. Another interesting idea is to illuminate the media with specific light, which will excite the superoxide radical anions. Since the distance between the oxygen atoms in the exited state is larger than in the ground state it will be easier to achieve the resonant diffusion. Moreover, the bond vibrations in the exited state possess larger magnitudes and thus they could lead to an additional increase of the diffusion coefficient.